\documentclass[%
 aip,
 jmp,%
 amsmath,amssymb,
 reprint,%
]{revtex4-1}

\usepackage{epsfig}
\usepackage{natbib}
\usepackage{color}
\usepackage{amsmath}
\usepackage{amssymb}
\usepackage{verbatim} 

\usepackage{graphicx}
\usepackage{dcolumn}
\usepackage{bm}

\usepackage{geometry} 
\newcommand{\mon}{\begin{displaymath}}
\newcommand{\moff}{\end{displaymath}}

\newcommand{\eon}{\begin{equation}}
\newcommand{\eoff}{\end{equation}}
\newcommand{\eaon}{\begin{eqnarray}}
\newcommand{\eaoff}{\end{eqnarray}}

\newcommand{\appropto}{\mathrel{\vcenter{
  \offinterlineskip\halign{\hfil$##$\cr
    \propto\cr\noalign{\kern2pt}\sim\cr\noalign{\kern-2pt}}}}}

\newcommand{\beq}{\begin{equation}}
\newcommand{\eeq}{\end{equation}}
\newcommand{\Bvec}{\mathbf{B}}
\newcommand{\vvec}{\mathbf{v}}
\newcommand{\vbgvec}{\mathbf{v}_{bg}}
\newcommand{\vv}{\mathbf{v}}

\newcommand{\vp}{v_\perp}

\newcommand{\eps}{\epsilon}

\newcommand{\lp}{\left(}
\newcommand{\rp}{\right)}

\begin{document}

\preprint{AIP/123-QED}

\title[Drift and Separation in Collisionality Gradients]{Drift and Separation in Collisionality Gradients}

\author{I. E. Ochs}
\affiliation{Department of Astrophysical Sciences, Princeton University, Princeton, New Jersey 08540}
\affiliation{Princeton Plasma Physics Laboratory, Princeton, New Jersey 08543}  

\author{J. M. Rax}
\affiliation{Departement de Physique, Universit\'{e} Paris XI-Ecole Polytechnique, LOA-ENSTA 9128 Palaiseau, France}

\author{R. Gueroult}
\affiliation{Laboratoire Plasma et Conversion d'Energie, CNRS, INPT, UPS, 31062 Toulouse, France}

\author{N. J. Fisch}
\affiliation{Department of Astrophysical Sciences, Princeton University, Princeton, New Jersey 08540}
\affiliation{Princeton Plasma Physics Laboratory, Princeton, New Jersey 08543}

\date{\today}

\begin{abstract}
We identify a single-particle drift resulting from collisional interactions with a background species, in the presence of a collisionality gradient and background net flow.
We analyze this drift in different limits, showing how it reduces to the well known impurity pinch for high-$Z_i$ impurities.
We find that in the low-temperature, singly-ionized limit, the magnitude of the drift becomes mass-dependent and energy-dependent. 
By solving for the resulting diffusion-advection motion, we propose a mass-separation scheme that takes advantage of this drift, and analyze the separative capability as a function of collisionally dissipated energy.
\end{abstract}

\maketitle

\twocolumngrid 
\section{Introduction}


It is often desirable to be able to separate elements based on mass.
In the production or reprocessing of nuclear fuel, for instance, uranium-235 must be separated from uranium-238; however, their identical chemical structure makes chemical separation impossible.
In another related application, the waste produced by nuclear reactors is extremely chemically inhomogenous, again making chemical separation difficult.
Fortunately, it is generally the case that the heavy elements present the greatest danger, while the lighter elements, such as Oxygen, Sodium, and Iron, are harmless but represent the bulk of the mass.
Thus useful separation on the basis of mass is possible.

The potential of rotating plasmas for discriminating elements based on atomic mass has long been recognized~\cite{Bonnevier1966,Lehnert1971}, and led to the development of plasma centrifuges for isotope separation~\cite{Krishnan1981,DelBosco1991,Grossman1991,yuferov2015physical}. However, only recently have plasmas been suggested for separating elements with large mass differences~\cite{Siciliano1993,Freeman2003,Zhiltsov2006,timofeev2014theory,Gueroult2014a,gueroult2015plasma}, such as needed for legacy waste disposal or spent fuel reprocessing. These emerging applications, focused more on throughput than fine-grained specificity, motivated the development of new mass filter concepts, including rotating plasma configurations~\cite{Ohkawa2002,fetterman2011magnetic,Gueroult2012a,gueroult2014double}, optical devices~\cite{Morozov2005,Bardakov2010}, crossed-field separators~\cite{Smirnov2013}, and filters based on differential magnetic drift~\cite{Timofeev2000} and gyro-radius effects~\cite{Babichev2014}.

It has long been known that a density gradient in a magnetized plasma can cause high-$Z_i$ impurity ions to be disproportionately drawn up the density gradient\cite{spitzer1952equations,taylor1961diffusion,braginskii1965transport,  hinton1976theory,hirshman1981neoclassical}.
This impurity pinch effect arises from the friction between the impurity ion and the bulk flow that results from the majority species diamagnetic drift, as well in inhomogeneities in the friction force about the gyro-orbit\cite{braginskii1965transport, taylor1961diffusion, hirshman1981neoclassical}.
Since this effect was seen as most relevant in fusion plasmas, models have generally assumed the presence of two thermal species at the same temperature, and focused on the $Z_i$-dependent effect of the equilibrium spatial distribution.
Thus mass-dependent and energy-dependent effects have remained relatively unexplored.

The paper is organized as follows.
In Section \ref{sec:drift}, we consider the motion of a single ion undergoing Langevin collisions with a background species, allowing for (a) a collisionality gradient, and (b) a bulk flow of the background species relative to the ion gyrocenter, and derive the resulting drift.
In Section \ref{sec:difbalance}, we consider the balance between this drift and collisional diffusion in determining the steady-state density gradient.
Applying this analysis to the case of ion-ion Coulomb collisions, we rederive the impurity pinch.
However, our approach also allows us to consider cases more relevant to the low-temperature plasmas desirable for mass filtration.
In particular, we show that even at constant $Z_i=1$, the drift-diffusion balance can lead to a mass-dependent and energy-dependent steady-state gradient.
In the subsequent sections, we show how this drift-diffusion balance can be exploited to separate ions of different mass, first in the context for a simple analytical model which ignores the ion heating mechanism (Section \ref{sec:separation}), and then for a more realistic heating model (Section \ref{sec:realisticHeating}).
Finally, we demonstrate how the energy expenditure per particle and separation power are intrinsically linked, and make estimates for the power consumption, separation, and throughput of the device.




\section{Derivation of the Drift}\label{sec:drift}


Consider an ion gyrating with speed $\vp$ in a $\hat{z}$-directed magnetic field, in the presence of a frictional collisionality gradient $\nabla \nu_s \parallel \hat{y}$.
Thus the collisionality locally takes the form
\beq
	\nu_s = \nu_0 + \nu' y.
\eeq
The differential equation that describes the ion motion, including both the Lorentz force and deterministic (frictional) collisions, is:
\begin{align}
	\frac{d \vv}{d t} &= \Omega \vv \times \hat{z} - \nu_s(y) (\vv - \vv_{bg}) \label{eq:ionDynamics}\\
	& = \Omega \vv \times \hat{z} - \lp \nu_0 + \nu' y \rp (\vv - \vv_{bg}),
\end{align}
where $\Omega$ is the gyrofrequency and $\vv_{bg}$ is the velocity of the background particles.
Because this is a single-particle equation, we do not include a thermal force; the subtleties involved with this choice are discussed in Appendix \ref{sec:thermalForce}.
Breaking into components, we have
\begin{align}
	\frac{d v_x}{d t} &= \Omega v_y - \lp \nu_0 + \nu' y \rp (v_x - v_{x,bg})\\
	\frac{d v_y}{d t} &= -\Omega v_x - \lp \nu_0 + \nu' y \rp (v_y  - v_{y,bg}).
\end{align}
Throughout the analysis, we will take $\Omega$, $\nu_0 \gg \nu' y$.
We will then consider the above equation in orders of $\varepsilon \equiv \nu' y/\Omega$.

To start, we shift to a 1D complex variable problem, letting
\begin{align}
	Z &= v_x^0 + iv_y^0\\
	z &= v_x^1 + iv_y^1,
\end{align}
where $Z = \mathcal{O}(1)$ and $z = \mathcal{O}(\varepsilon)$.
The particle positions can then be found via
\begin{equation}
	x + iy = x_0 + iy_0 + \int_0^t Z(\tau) d\tau + \int_0^t z(\tau) d\tau.
\end{equation}
Start by solving the zeroth-order equation:
\begin{align}
	\frac{dZ}{dt} &= \frac{d v_x^0}{d t} + i\frac{d v_y^0}{d t} = -(i \Omega + \nu_0) Z + \nu_0 z_{bg},
\end{align}
where $z_{bg}$ is the complex representation of $\vv_{bg}$.
The general solution to the equation
\beq
	\frac{dx}{dt} = -\omega x + S
\eeq
is
\beq
	x = x_0 e^{-\omega t} + e^{-\omega t} \int_0^t e^{\omega \tau} S(\tau) d\tau .
\eeq
Thus we have:
\begin{align}
	Z(t) &= e^{-(i \Omega + \nu_0) t} \lp C + \int_0^t \nu_0 z_{bg}e^{(i \Omega + \nu_0) \tau} d\tau\rp \\
	&= Z_{0} e^{-(i \Omega + \nu_0) t} + \frac{\nu_0 z_{bg}}{i \Omega + \nu_0}, \label{eq:zerothOrderComplex}
\end{align}
where we have combined the arbitrary constant $C$ with the $t=0$ contribution from the integral into $Z_0$.

The second term, corresponding to the background flow drift, is given by:
\begin{align}
	v_{d,\text{flow}} &= \frac{\nu_0 (v_{x,bg} + i v_{y,bg})}{\Omega^2 + \nu_0^2} \lp \nu_0 - i \Omega \rp\\
	&=  \frac{\nu_0 }{\Omega^2 + \nu_0^2} \left[ \nu_0 (v_{x,bg} + iv_{y,bg}) - \Omega (-v_{y,bg} + i v_{x,bg}) \right]
\end{align}
In terms of vectors,
\begin{align}
	\vv_{d,\text{flow}} &=   \frac{\Omega \nu_0 }{\Omega^2 + \nu_0^2} \left(  \vv_{bg} \times \hat{b} +\frac{\nu_0}{\Omega} \vv_{bg}   \right) \label{eq:vdflow}
\end{align}

Although we solved for this drift in the zeroth-order equation, in general we will be interested in cases where $v_{bg} \ll  \vp$, i.e. $|z_{bg} /  Z_0| = \mathcal{O} (\varepsilon)$.
Thus we will ignore terms of order $\nu' z_{bg}$ in our first-order equation, since these are $\mathcal{O} (\varepsilon^2)$.
Thus our first-order equation becomes:
\begin{align}
	\frac{dz}{dt} &= \frac{d v_x^1}{d t} + i\frac{d v_y^1}{d t} = -(i \Omega + \nu_0) z - \nu' y^0 Z,
\end{align}
which has the standard solution:
\beq
	z = e^{-(i \Omega + \nu_0) t} \lp z_0  - \int_0^t e^{(i \Omega + \nu_0) \tau} \nu' y^0(\tau) Z(\tau) d\tau \label{eq:zIntegral}\rp.
\eeq

First we need to calculate $y$.
Because we are interested in cases where $|z_{bg} /  Z_0 | = \mathcal{O} (\varepsilon)$, we will ignore the second term (the drift term) in Eq. (\ref{eq:zerothOrderComplex}) when calculating the product with the first-order term $\nu'$; i.e., we take
\beq
	Z(t) \approx Z_{0} e^{-(i \Omega + \nu_0) t}.
\eeq
We then have, noting $Z_0 \approx v_{x0} + i v_{y0}$,
\begin{align}
	y^0  &= y_0^0 + \text{Im}\left[  \int_0^t Z(\tau) d\tau \right]\\
	&\approx  \lp \frac{e^{-\nu_0 t}}{\Omega^2 + \nu_0^2} \rp [ ( \nu_0 v_{x0} + \Omega v_{y0}) \sin \Omega t \notag\\
	& \qquad \qquad \qquad \; + ( \Omega v_{x0} -\nu_0 v_{y0}) \cos \Omega t ] \notag \\
	& \qquad \qquad + y_0^0 - \lp \frac{\Omega v_{x0} -\nu_0 v_{y0}}{\Omega^2 + \nu_0^2} \rp.
\end{align}
Now, we are really interested in the gyro-drift of a particle orbiting the origin; the last term simply accounts for the shifted rotation center if the particle starts at the origin.
Thus we take $y_0^0$ to cancel this term, leaving only the oscillating terms.

Meanwhile,
\begin{align}
	e^{(i \Omega + \nu_0) \tau} Z(\tau)  \approx v_{x0} + i v_{y0}.
\end{align}
Averaging over an isotropic perpendicular velocity distribution, take $\langle v_{x0} v_{y0} \rangle =0$, and $\langle v_{x0} v_{x0} \rangle = \langle v_{y0} v_{y0} \rangle = v_{\perp 0}^2/2$, so that we have
\begin{align}
	\langle e^{(i \Omega + \nu_0) t} \nu' y_0 Z \rangle &= \frac{1}{2} \lp \frac{v_{\perp0}^2 \nu'}{\Omega^2 + \nu_0^2} \rp e^{-\nu_0 t} [ ( \nu_0  + \Omega i ) \sin \Omega t  \notag\\
	& \qquad \qquad \qquad \qquad \qquad + ( \Omega -\nu_0 i ) \cos \Omega t ]\\
	&= \frac{1}{2} \lp \frac{v_{\perp0}^2 \nu'}{\Omega^2 + \nu_0^2} \rp \left[ \Omega - \nu_0 i  \right] e^{(i\Omega-\nu_0) t} \label{eq:zIntegrand}
\end{align}
Then, taking $z_0$ so as to cancel the $t=0$ contribution from the integral in Eq. (\ref{eq:zIntegral}), we can integrate Eq. (\ref{eq:zIntegrand}) to find:
\begin{align}
	z(t) 	&= \frac{1}{2} \lp \frac{v_{\perp0}^2 \nu'}{\Omega^2 + \nu_0^2}\rp e^{-2\nu_0 t} \lp \frac{2 \Omega \nu_0}{\Omega^2 + \nu_0^2}  + \frac{\Omega^2 - \nu_0^2}{\Omega^2 + \nu_0^2} i \rp  
\end{align}

Now simply from slowing, we have $\vp = v_{\perp0} e^{-\nu_0 t}$, so
\begin{align}
	v_{d,c} = z(t) = \frac{1}{2} \lp \frac{\vp ^2 \nu'}{\Omega^2 + \nu_0^2}\rp \lp \frac{2 \Omega \nu_0}{\Omega^2 + \nu_0^2} \hat{x} + \frac{\Omega^2 - \nu_0^2}{\Omega^2 + \nu_0^2} \hat{y} \rp.
\end{align}

Combining the two drifts by adding back the second term in Eq. (\ref{eq:zerothOrderComplex}), which become Eq. (\ref{eq:vdflow}), we have:
\begin{align}
	\vv_d  &=  \frac{1}{2} \lp \frac{\vp ^2}{\Omega^2 + \nu_0^2}\rp \lp \frac{\Omega^2 - \nu_0^2}{\Omega^2 + \nu_0^2} \nabla \nu + \frac{2 \Omega \nu_0}{\Omega^2 + \nu_0^2} \nabla \nu \times \hat{b}  \rp  \notag\\
	& \qquad \qquad +\frac{\Omega \nu_0 }{\Omega^2 + \nu_0^2} \left(  \vv_{bg} \times \hat{b} +\frac{\nu_0}{\Omega} \vv_{bg}   \right) 
\end{align}
As $\nu_0/\Omega \rightarrow 0$, we have
\beq
	\vvec_d = \frac{1}{2} \rho^2 \nabla \nu_s  + \frac{\langle \nu_s \rangle}{\Omega} \vbgvec \times \hat{b}  + \mathcal{O}\lp \frac{\nu_0}{\Omega} \rp . \label{eq:heuristic_drift}
\eeq
Each of these terms has a clear physical interpretation.
The second term simply represents the $F \times B$ (generalized Hall) drift due to momentum transfer from the background flow.
It is primarily this second term that gives rise to the ``impurity pinch'' experienced by high-$Z_i$ ions in tokamaks, as a result of the diamagnetic poloidal flow friction.
The first term, meanwhile, represents the $F \times B$ drift that arises from the difference in collisional friction between the upper and lower halves of the orbit due to the collisional inhomogeniety.
These heuristics are discussed in more detail in Appendix \ref{sec:heuristic_drift}.

\section{Balance with Diffusive Terms and Steady State}\label{sec:difbalance}
The drift we have discussed thus far results from the deterministic drag term in the Langevin equation.
Now we briefly review the impact of the diffusive terms.
Because the diffusion is isotropic over a gyro-orbit, the diffusion in the drift will manifest itself as diffusion in guiding-center position, which is given by the classical expression
\beq
	D_\perp = \frac{1}{2} \frac{\vp^2 \nu_D}{\Omega^2+\nu_D^2} \sim \frac{1}{2}\rho^2 \nu_D, \label{eq:dperp}
\eeq
where we have taken care to distinguish $\nu_D$, the diffusive collision term, from $\nu_s$, the slowing collision term.

In diffusion-advection processes, the steady-state gradient depends on the ratio of advection to diffusion.
Since $v_d \sim \nu_s$, while $D \sim \nu_D$, this leads us to define the important parameter
\beq
	\chi \equiv \nu_s/\nu_D,
\eeq 
the ratio between the slowing and diffusive collision terms.
Thus, as the slowing down terms become more significant, $\chi$ increases, and the steady-state ion gradient becomes steeper.

It should be noted that the diffusion process we are describing here is not directly comparable to the situations considered by Braginskii.
Whereas the Braginskii equations examine transport coefficients resulting from fluxes averaged over a Maxwellian distribution, here we examine the fluxes for test particles at a certain fixed energy.
This allows us to consider more general distributions--in particular, sharply-peaked, high-energy distributions which could be desirable for separations.

\subsection{Mass and temperature dependence of $\chi$}

Hot, heavy particles will tend to be more susceptible to slowing than light particles. 
Thus we expect different gradients in steady-state as a function of mass and energy.
To see this tendency, our first task is to be explicit in what we mean by $\nu_D$.
We care about diffusion along a single dimension  perpendicular to the particle orbit.
In the case of ion-cyclotron heating, we will have $\eps_\perp \gg \eps_\parallel$, so 
\beq
	\nu_{D,ic} = \frac{1}{2} \lp \frac{1}{2} \nu_\perp + \nu_\parallel \rp = \frac{1}{4} \nu_\perp + \frac{1}{2} \nu_\parallel. \label{eq:nu_D_ic}
\eeq
Here, the first factor of 1/2 comes from considering diffusion along a single dimension perpendicular to the magnetic field. 
The factor of 1/2 before $\nu_\perp$ comes from the fact that only diffusion along one of the two dimensions perpendicular to the particle velocity will result in diffusion perpendicular to the magnetic field, since one of the perpendicular dimensions lies parallel to the magnetic field.

Meanwhile, in the case of isotropically distributed ion velocities, we will have
\beq
	\nu_{D,iso} = \frac{1}{3} \lp \nu_\perp + \nu_\parallel \rp.
\eeq
This would be the proper form, for instance, for thermal ions or fusion-born $\alpha$ particles.

\subsubsection{Collisions with fast ions and electrons}

With our diffusion collision frequency in hand, it is now possible to consider limiting cases of $\chi$ in several cases.
First we consider the low-velocity limit of the collision coefficients, in which $\frac{\eps_{ij}}{x_{ij}} \ll 1$,
where $x_{ij} \equiv m_i / m_j$ is the mass ratio with the buffer species, and $\eps_{ij} = \eps_{\perp i}/T_j$ is the ratio of the perpendicular ion energy to the buffer temperature.
In this limit, $\nu_\perp = 2\nu_\parallel$, so $\nu_{D,ic} = \nu_{D,iso}$.
Then, using the formulary frequencies:
\begin{align}
	\chi_{is} &= \frac{\eps_{ij}}{\lp 1 + \frac{1}{x_{ij}} \rp^{1/2}} \rightarrow \begin{cases}
	\eps_{ij} x_{ij}^{1/2} &\text{ if } x_{ij} \ll 1\\
	\eps_{ij} &\text{ if } x_{ij} \gg 1.
	\end{cases}
\end{align}
In the important case of heavy thermal ions, where $\eps_{ij} = T_{ij}$, we have $\chi_{is} = 1$.

When considering ion-electron collisions, we will almost always be in the low-velocity limit.
Then
\beq
	\chi_{es} = \eps_{ie}.
\eeq
Thus differentially heating ions will result in different drift vs diffusion rates on electrons proportional to the energy.

\subsubsection{Collisions with slow ions}

When $\frac{\eps_{ij}}{x_{ij}} \gg 1$, we find slightly different results for isotropic vs ion-cyclotron heated ions.
First consider isotropically-directed ions.
Then we have
\begin{align}
	\chi_{if,iso} &= \frac{\lp \mu_i^{-1} + \mu_j^{-1} \rp \mu_i^{1/2} \eps^{-3/2}}{\tfrac{2}{3} \mu_i^{-1/2} \eps^{-3/2} + \tfrac{1}{3}\mu_i^{1/2}\mu_j^{-1} \eps^{-5/2} T_j}\\
	&= \frac{1 + x_{ij}  }{\tfrac{2}{3}   + \frac{1}{3} \tfrac{x_{ij}}{\eps_{ij}}}\\
	& \approx \frac{3}{2} (1 + x_{ij}).
\end{align}

Meanwhile, for ion-cyclotron heating:
\beq
	\chi_{if,ic}  =  \frac{1 + x_{ij}}{\tfrac{1}{2} +  \tfrac{1}{2}\tfrac{x_{ij}}{\eps_{ij}}} \approx 2(1+x_{ij}).
\eeq
Thus fast, heavy ions will experience strong collisional drift relative to diffusion.

\subsection{Special case: background ion gradient with diamagnetic flow}

In the special case of ions diamagnetic flow due to a density gradient in the $x$ direction, we have
\begin{align}
	\vv_{bg} &= -\frac{T \nabla \log n_{bg} \times \hat{b}}{Z_{bg} e B}.
\end{align}
Thus, since $\nu_s \propto n_{bg}$, we have
\begin{align}
	\vv_d &= \frac{1}{2} \rho^2 \langle \nu_s \rangle \lp \nabla \log n_{bg} - \frac{2 \Omega}{v_\perp^2} \lp \frac{T \nabla \log n_{bg} \times \hat{b}}{Z_{bg} e B} \rp \times \hat{b}\rp\\
	&= \frac{1}{2} \rho^2 \langle \nu_s \rangle \nabla \log n_{bg} \lp  1 + \frac{2 (Z_i e B)}{m_i v_\perp^2} \frac{T }{Z_{bg} e B} \rp\\
	&= \frac{1}{2} \rho^2 \langle \nu_s \rangle \nabla \log n_{bg} \lp  1 + \frac{Z_i}{Z_{bg} } \frac{1}{\eps_{ij}}\rp.
\end{align}

From this result, we can recover the well-known high-$Z_i$ impurity pinch\cite{taylor1961diffusion}.
The flux is given by
\beq
	\Gamma = v_d n_I - \frac{d}{dx} \lp n_I D \rp.
\eeq
Now take $\nu_s = \tilde{\nu}_s n_{bg}$, $\nu_D = \tilde{\nu}_D n_{bg}$, so that the dependence on density becomes more apparent.
Then, plugging in for $v_d$ and $D$ in the limit $\nu \ll \Omega$, we have
\begin{align}
	\Gamma &= \frac{1}{2} \rho^2 \tilde{\nu}_s  \lp  1 + \frac{Z_i}{Z_{bg} } \frac{1}{\eps_{ij}}\rp n_I \frac{d}{dx}  n_{bg}  - \frac{1}{2} \rho^2 \tilde{\nu}_D \frac{d}{dx} \lp n_{bg} n_I  \rp.\\
	& = \frac{1}{2} \rho^2  \tilde{\nu}_D \left[ (\chi - 1) n_I \frac{dn_{bg}}{dx} +  \frac{Z_i}{Z_{bg} } \frac{\chi}{\eps_{ij}} n_I \frac{dn_{bg}}{dx} - n_{bg} \frac{dn_I}{dx} \right]
\end{align}
Now for the case of thermal, heavy ions, $\chi = \eps_{ij} = 1$, so
\begin{align}
	\Gamma &= \frac{1}{2} \rho^2 \tilde{\nu}_D \left[ \frac{Z_i}{Z_{bg} } n_I \frac{dn_{bg}}{dx} - n_{bg} \frac{dn_I}{dx} \right]
\end{align}
If we set $\Gamma = 0$ for steady state, and take $\eps_{ij} = 1$ (i.e. consider thermal particles), then we find
\beq
	\frac{n_{bg}^{Z_I/Z_{bg}}}{n_I} = \text{const},
\eeq
which is the impurity pinch result.
The full impurity pinch, including the reversal in steep temperature gradients, could be derived similarly by including the temperature-dependent terms in both the diamagnetic flow velocity and the collisionality.

This derivation relied on the assumption that $\tilde{\nu}_s \neq \tilde{\nu}_D$, which is not true in general.
As we show next, the non-cancellation of these terms in general can result in interesting energy-dependent and mass-dependent separative effects, even between species with equal $Z$.

\section{Mass-based separation scheme}\label{sec:separation}
In this section, we explore how the energy- and mass-dependence of the diffusion/advection ratio could be used to separate particles of different masses.

\subsection{Diffusion model}

Consider a system extending over a distance $0< x < L_s$ perpendicular to a magnetic field, with a source at $x = L_-$ and sinks at $x=0$ and $x = L_s$ given by $S(x) = -c\delta (x) + \delta (x-L_-) - (1-c)\delta (x - L_s)$.
Our steady-state diffusion equation is then given by
\beq
	0 = v_d(x) f - \frac{d}{dx} (D(x) f) - U(x),
\eeq
where 
\beq
	U(x) = \int_0^x S(s) ds =  -c + H(x-L_-) 
\eeq
is the integrated flux, and $H(x)$ is the heaviside function.
Evaluating the derivative using the product rule and rearranging, we have
\beq
	\frac{df}{dx} = \lp \frac{v_d(x)}{D(x)} - \frac{1}{D(x)}\frac{d D(x)}{dx} \rp f - \frac{U(x)}{D(x)}. \label{eq:diff_adv}
\eeq

\begin{figure}[b] 
	\center{\includegraphics[width=\linewidth]{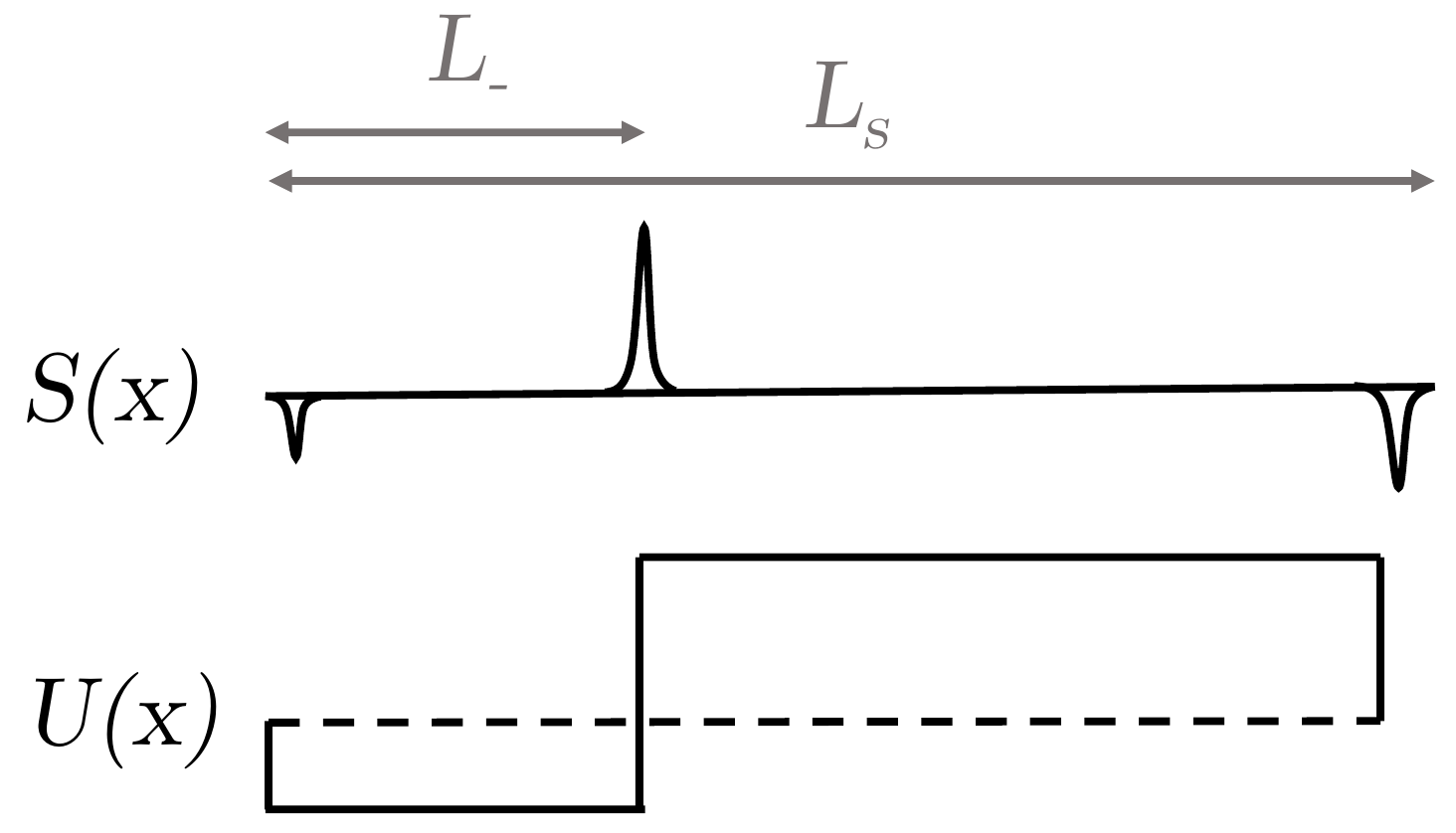}}
	\caption{Diffusion problem setup, showing source and sink terms $S(x)$ and integrated source flux $U(x)$ for steady-state diffusion.}
	\label{fig:diff_flux}
\end{figure}


Up until now, the functional dependencies of $D$ and $v_d$ have been treated in full generality.
Now consider a system in which $T$ is constant, but the density varies as $n = n_0 e^{x/L_c}$. 
Also assume that the particle is maintained at a constant energy $\eps_{ij}$.
Then, plugging in for $v_d$ and $D$, we have
\beq
	\frac{df}{dx} = \frac{\eta - 1}{L_c} f - J e^{-x/L_c} \lp -c + H(x - L_-) \rp ,
\eeq
where 
\beq
	\eta = \lp  1 + \frac{Z_i}{Z_{bg} } \frac{1}{\eps_{ij}}\rp \chi
\eeq
and
\beq
	J = \frac{2}{\rho^2 \tilde{\nu}_D n_0}. \label{eq:J}
\eeq
Now, the solution to this equation is
\beq
	f(x) = e^{-\int_{0}^{x} P(s) ds} \lp \int_{0}^{x} e^{\int_{0}^{y} P(s) ds} Q(y) dy + d \rp.
\eeq
where 
\beq
	P(x) = -\frac{\eta - 1}{L_c}, \quad Q(x) = -J e^{-x/L_c} \lp -c + H(x - L_-) \rp. \label{eq:pq}
\eeq
Then
\beq
	\mp \int_{0}^{x} P(s) ds = \pm (\eta - 1) \frac{x}{L_c}.
\eeq
Thus
\begin{align}
	f(x) &= \frac{J L_c}{\eta}  e^{(\eta-1) \frac{x}{L_c}} \biggl[ \lp 1- e^{-\eta \frac{x}{L_c}} \rp c \notag \\
	&\qquad - \lp e^{-\eta \frac{L_-}{L_c}}- e^{-\eta \frac{x}{L_c}}\rp H(x - L_-)    + d \biggr] \label{eq:f}
\end{align}

Now we can apply our boundary conditions.
Since $f(0) =0$, we see easily that $d=0$.
Then, solving for $c$ by setting $f(L_s) = 0$, we find
\beq
	c = \frac{e^{\eta \frac{(L_s-L_-)}{L_c}}- 1} {e^{\eta \frac{L_s}{L_c}}- 1}.
\eeq
The relative fluxes $F_\mp$ of particles leaving from the left ($-$) and right ($+$) sides of the device are thus
\begin{align}
	F_-&= c = \frac{e^{\eta \frac{(L_s-L_-)}{L_c}}- 1} {e^{\eta \frac{L_s}{L_c}}- 1}\\
	F_+&= 1-c = \frac{1-e^{-\eta \frac{L_-}{L_c}}} {1-e^{-\eta\frac{L_s}{L_c}}}. \label{eq:Fp}
\end{align}

Meanwhile, the maximum density of non-buffer particles in the device scales as
\beq
	n_{\max} = \frac{2 L_c}{\eta \rho^2 \tilde{\nu}_D n_0} \lp \frac{\dot{N}}{L_y L_z} \rp \max \tilde{f}(x),
\eeq
where $\dot{N}$ is the total particle flow through the mass filter, $L_y$ and $L_z$ are the two remaining device dimensions, and $\tilde{f}(x) = f(x)\eta/JL_c$ is the normalized density function.
$\tilde{f}(x)$ generally has its maximum at $x=L_-$, where
\begin{align}
	\tilde{f}(L_-) &= \frac{\lp e^{(\eta-1) L_-/L_c}-1\rp \lp e^{\eta(L_s-L_-)/L_c}-1\rp}{e^{\eta L_s/L_c}-1}\\
	&\approx (\eta-1) \frac{ L_-}{L_c}F_-
\end{align}
Thus,
\beq
	\frac{\dot{N}}{n_{\max}} \approx L_y L_z \frac{\rho^2 \nu_{D0}}{2 L_- F_-} \frac{\eta}{\eta-1}
\eeq
For practical applications, there should be several gyro-radii between the injection point and the periphery, so take $L_-/\rho \approx 3$. 
We also want to maintain ion gyro-drift motion even at the region of highest density, so take $\nu_{D0}/\Omega \approx 3$.
These could be approximately satisfied, for instance, by $\mu_{bg} = 1$, $\mu_i = 100$, $B = 2000$ G, $T_{bg} = 1$ eV, $\eps_{ij} = 5$, and $L_- = 3$ cm. 
Then, recalling $\rho = v_{thi}/\Omega$,
\beq
	\frac{\dot{N}}{n_{\max}} \approx \frac{1}{10} L_y L_z \frac{v_{thi}}{2 F_-} \frac{\eta}{\eta-1}
\eeq
For a heavy ion mass of 100 amu and temperature of 5 eV, we have $v_{thi} \approx 2 \times 10^5$ cm/s. 
Assuming a collection area of $L_y L_z = 300$ cm$^2$ and $F_- \approx 0.3$, we then have $\dot{N}/n_{\max} \approx 10^7$.
Thus, for a relatively small core heavy ion density of $n_h=10^{12}$ cm$^{-3}$, the device would process $10^{19}$ ions/second, or around a milligram per second.

\subsection{Power dissipation}

Now let us assume that one of the species (the heavy minority) is heated by ICRF, in order to keep $\chi$ (and thus $\eta$) much larger for this species.
The power dissipation is approximately given by
\beq
	P_{dis} = \eps_\perp \, \nu_\eps = \eps_\perp \lp 2 \nu_s - \tfrac{1}{2}\nu_\perp - \nu_\parallel \rp,
\eeq
where the factor of $1/2$ in front of $\nu_\perp$ arises because energy added parallel to the magnetic field (one of the directions perpendicular to $\vv_\perp$) does not help to maintain the perpendicular energy.
Using Eq. (\ref{eq:nu_D_ic}), this can be written
\beq
	P_{dis} = 2 \eps_\perp \nu_D \lp \chi_h - 1 \rp,
\eeq
where the $h$ subscript reminds us that we are only heating the heavy species.
Thus, recalling the density-independent quantity $\tilde{\nu}_s \equiv \nu_s/n$, the power dissipated throughout the device is given by
\begin{align}
	P_{dis,\text{tot}} &= \int_0^{L_s} P_{dis} (x) f(x) dx \label{eq:pdis}\\
	&= 2 \eps_\perp \tilde{\nu}_D  \lp \chi_h - 1 \rp \int_0^{L_s} n(x) f(x) dx.
\end{align}
Plugging in Eqs. (\ref{eq:f}), (\ref{eq:J}), and (\ref{eq:Fp}) and carrying out the integral, we find
\begin{align}
	P_{dis,\text{tot}} &= 4 \eps_\perp \frac{\chi_h-1}{\eta_h} \frac{L_c}{\rho^2} \left( L_s F_+ - L_- \right).
\end{align}
Note that, when $F_+ = L_-/L_s$, i.e. when the random walk is unbiased, there is no power dissipation.
Because this calculation shows the energy dissipated per unit time for a system with a feed rate of one particle per unit time (since $F_+ + F_- = 1$), we can interpret this dissipated power as the mean energy dissipated per particle traversing the system. 

Since for a slow (but relatively energetic), singly-ionized heavy particle, $\eta_h -1 = \chi_h = \eps_{ij}$, we can write
\begin{align}
	P_{dis,\text{tot}} &= 4 T \frac{(\eta_h - 1)(\eta_h-2)}{\eta_h} \frac{L_c}{\rho^2} \left( L_s F_+ - L_- \right). \label{eq:pdis_simp}
\end{align}
Meanwhile, light particles are not heated, so they dissipate no energy, and have $\eta_l = 2$.

\begin{figure}[b] 
	\center{\includegraphics[width=\linewidth]{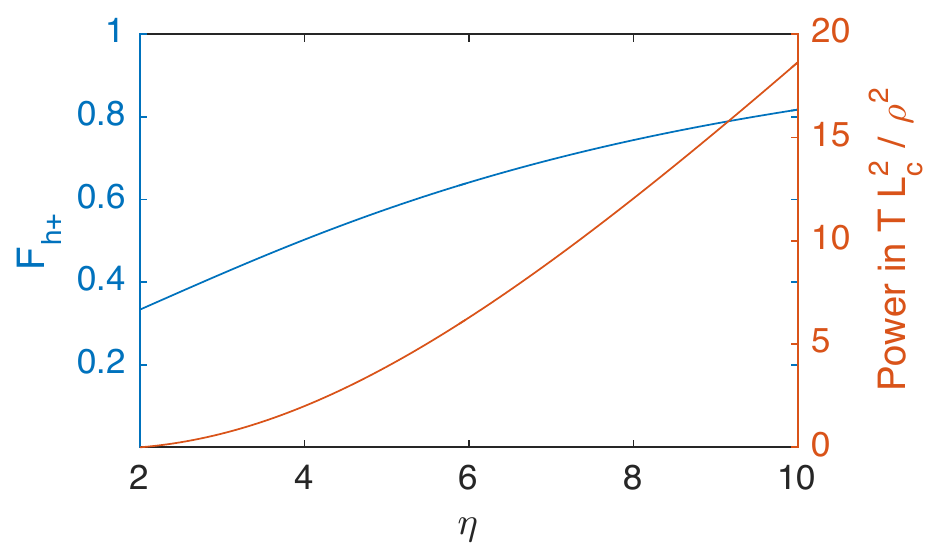}}
	\caption{Fraction of particles leaving the right side of the device and energy dissipated per particle (in units of $T L_c^2/\rho^2$), as a function of $\eta$.
	Here we take $L_s = L_c$, $L_- = 0.17 L_c$.
	Particles are assumed to be slow but energetic relative to the much lighter buffer gas.
	At $\eta_h = 6$, around two thirds of the heavy particles exit to the right, and two thirds of the light particles ($\eta_l=2$) exit to the left, while dissipating $6 T L_c^2/\rho^2$, around 600 eV for a 1eV buffer gas, per particle.}
	\label{fig:sep_efficiency}
\end{figure}

Figure \ref{fig:sep_efficiency} shows the evolution of $F_{h+}$ and $P_{dis,\text{tot}}$ with $\eta_h$ for $L_s = L_c$ and $L_m = 0.17 L_c$.
At $\eta_h = 6$, approximately two thirds of the heavy elements exit on the right side of the device, while approximately two thirds of the light elements ($\eta_l = 2$) exit on the left side, representing a twofold enrichment of the heavy element in the output stream.
The dissipated power to achieve this separation is $6.3 T L_c^2/\rho^2$, which for a 1eV plasma with $L_c/\rho \approx 10$ would be 600 eV per heavy particle.

%

\section{A more realistic heating model}\label{sec:realisticHeating}

In the previous section we assumed that the energy of each species was spatially uniform.
In realistic ICRF heating scenarios, the energy which the particle gains from the heating will be dissipated through collisions; thus the energy will generally be lower in areas of high density.
To get the energy as a function of position, assume the particle is in force balance.
Then:
\beq
	F + \lp - \nu_s  + \frac{1}{4} \nu_\perp + \frac{1}{2} \nu_\parallel \rp v_\perp = 0,
\eeq	
where $F$ is a forcing term representing the ICRF.
This gives
\begin{align}
	\eps_\perp = \lp -\frac{B}{2A} \lp 1 + \sqrt{1 - 4 \frac{AC}{B^2}} \rp \rp^2
\end{align}
where
\begin{align}
	A &= \frac{\mu_j^{1/2}}{\mu_i} \lp 1 + \frac{\mu_j}{\mu_i} \rp^{-1/2} T_i^{-3/2}\\
	B &= -10.7 \frac{F \mu_i^{1/2} }{\lambda n_{bg}(x)}\\
	C &= -\frac{\mu_j^{1/2}}{\mu_i} T_i^{-1/2}
\end{align}
So, plugging in:
\begin{align}
	\eps_\perp &= \eps_0 \lp 1 + \sqrt{1+\eps_0^{-1} \lp 1+\tfrac{\mu_j}{\mu_i}\rp^{1/2}  T_i} \rp^2 \label{eq:eps_heat}
\end{align}
where
\beq
	\eps_0 \equiv 28 \frac{F^2 T_i^3 \mu_i^2 \lp 1 + \tfrac{\mu_i}{\mu_j} \rp}{\lambda^2 n_{bg}(x)^2}. \label{eq:eps0}
\eeq
Thus $\eps_\perp$ scales as $n_{bg}^{-2}$ until it reaches a minimum of $\eps_\perp = T_i$, where it saturates.
This means that $\eta$ also scales as $n_{bg}^{-2}$ before saturating at $\eta = 2$.

\subsection{Numerical results for realistic heating} \label{sec:numerical}

We can use Eq. (\ref{eq:eps_heat}) to calculate the drift and diffusion terms in Eq. (\ref{eq:diff_adv}) and Eq. (\ref{eq:pdis}).
By numerically solving these equations consistently with the boundary conditions, we can then find both the separation and the energy dissipation per particle for a given device.

As an example, we consider a 100-amu particle in a separation device, extending over $0 < x < L_s - L_-$, with injection at $x = L_-$, where the density of 1-amu background particles is given by
\beq
	n_{bg}(x) = n_0 (1+(x-L_-)/L_n). \label{eq:dev_heat}
\eeq
We take $B = 10^4$ G, $T_i = 0.7$ eV, and varied the forcing term $F$ from $10^7$ to $10^{10}$ cm/s$^2$.
Recalling from Eq. (\ref{eq:pdis_simp}) that better efficiency is generally achieved by making $L_s/ \rho_i$ as small as feasibly possible, we take $L_n = 12\rho_{i0}$, $L_- = 2 \rho_{i0}$, and $L_s = 10\rho_{i0}$, where $\rho_{i0}$ is the equilibrium gyroradius at the injection point.
Thus the device size was larger for greater forcing $F$, and smaller for lower $F$.

\begin{figure}[t] 
	\center{\includegraphics[width=\linewidth]{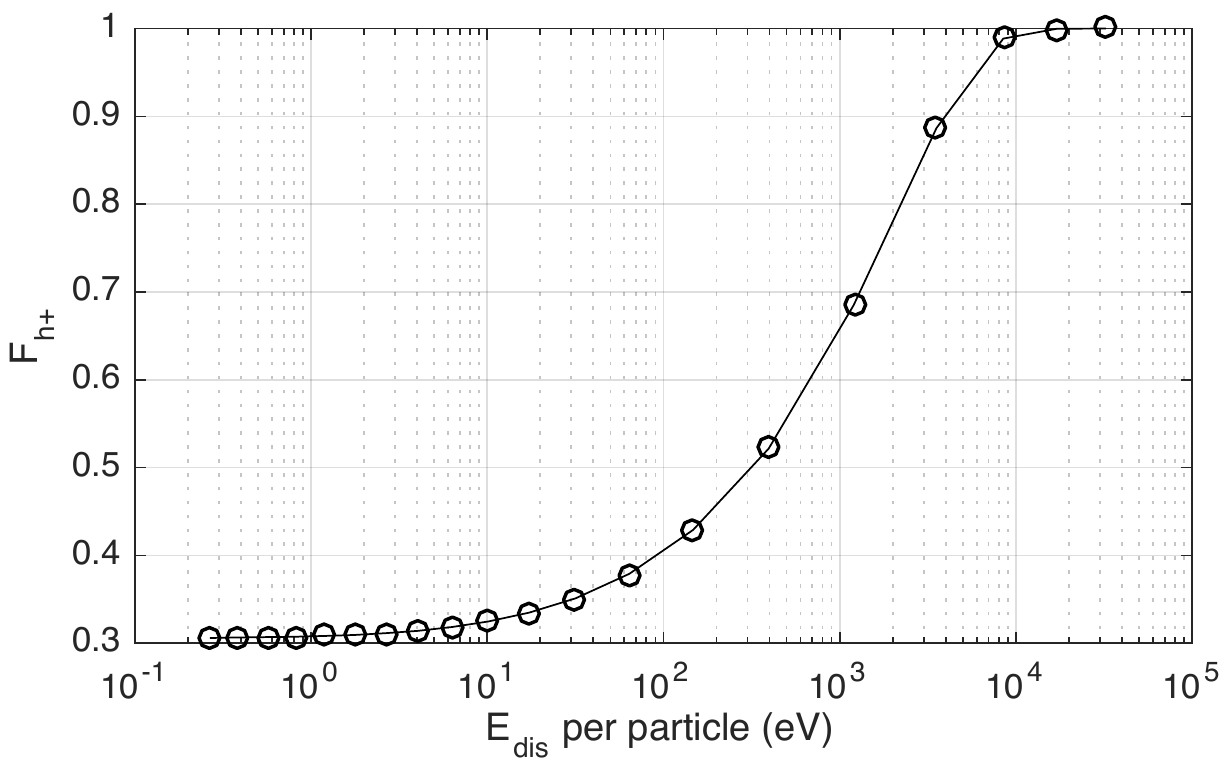}}
	\caption{Fraction of particles leaving the right side of the device as a function of energy dissipated per particle for the device in Section \ref{sec:numerical}.
	Thus $B = 10^4$ G, $T_i = 0.7$ eV, $L_n = 12\rho_{i0}$, $L_- = 2 \rho_{i0}$, and $L_s = 10\rho_{i0}$, and the forcing term $F$ was varied from $10^7$ to $10^{10}$ cm/s$^2$.}
	\label{fig:sep_heat}
\end{figure}

Results for the fraction of particles exiting at the right side of the device as a function of power dissipated are shown in Figure \ref{fig:sep_heat}.
The tradeoff between energy expenditure and separation is clear from the figure.
As in Figure \ref{fig:sep_efficiency}, 60\% / 30\% separation occurs in the neighborhood of 1 keV.
However, it is apparent that by pushing to higher power, arbitrarily high separation is possible.
For instance, at 85 keV dissipated, 98.9\% of heated particles exit on the right side of the device.
Thus, depending on the separation factor required, a given device could be tuned to be low-separation, low-energy, as for waste filtration, or high-separation, high-energy, as for isotope separation.

\section{Discussion and Conclusions}

Starting from a single-particle model and adding Langevin collisions with an inhomogenous, flowing background, we have derived expressions for the associated gyrocenter drifts and diffusion.
We find that the velocity-space Langevin drag term gives rise to the gyrocenter drift, and the velocity-space Langevin diffusion term gives rise to the gyrocenter diffusion.
This general model is applicable to various scenarios, including collisions with a gradient of (diamagnetically-drifting) charged particles, and collisions with an inhomogenous population of neutrals.

In focusing on ion-ion collisions, we first recovered the well-known impurity pinch \cite{spitzer1952equations,taylor1961diffusion,braginskii1965transport,  hinton1976theory,hirshman1981neoclassical}, known to cause high-$Z$ impurities to congregate in the tokamak core.
We then focused on singly-ionized populations, finding mass- and energy-dependent drifts.
Exploiting these energy dependencies, we outlined a novel mass separation scheme, and analyzed the energy/separation tradeoff, finding that substantial separation could be accomplished at around one keV per particle.
Importantly, the same device could achieve greater separation at the cost of increasing the dissipated power, raising the possibility that the general scheme could be used for either waste reprocessing or isotope separation.

The separation energy of a few keV per particle is much greater than less targeted bulk plasma separation methods,\cite{gueroult2015plasma,fetterman2011magnetic,gueroult2014double} which claim a plasma energy cost of 1.5 GJ/kg to separate aluminum, i.e. 400 eV/particle. 
However they are on the same order as other ICR schemes\cite{timofeev2007plasma,potanin2008extraction,timofeev2009plasma,timofeev2014theory, dawson1976isotope}.
In contrast to the proposed scheme, these existing schemes (a) rely on very low densities to avoid collisional dissipation as the ions reach energies on the keV scale, and (b) require the heated species to be recovered by embedding in a solid physical medium. 
The first constraint is detrimental because it requires a high-vacuum system, and also limits the throughput.
It also results in wall degradation, since ions strike the wall at energies in excess of 1 keV.
The second constraint further limits the throuput, by requiring a stage of operation where waste is physically scraped off the embedding medium.
In contrast, the proposed device both works at higher density, and allows for either species to be heated and extracted by any method at opposite ends of the device, while maintaining the particle energy below $\sim 50$ eV.

\begin{acknowledgments}
This work was performed under U.S.~DOE contract DE-AC02-09CH11466. 
One author (IEO) also acknowledges the support of the DOE Computational Science Graduate Fellowship.
\end{acknowledgments}

\appendix

\section{Heuristic Description of the Gradient Drift} \label{sec:heuristic_drift}

Consider an ion interacting with a background species via collisions.
Drifts arise from both (a) inhomogeneities in collisions around the gyro-orbit, due to gradients in temperature or density of the background, and (b) net flow velocity of the background with respect to the gyrocenter rest frame. 
In this section, we heuristically derive these two drifts.

\begin{figure}[b] 
	\center{\includegraphics[width=\linewidth]{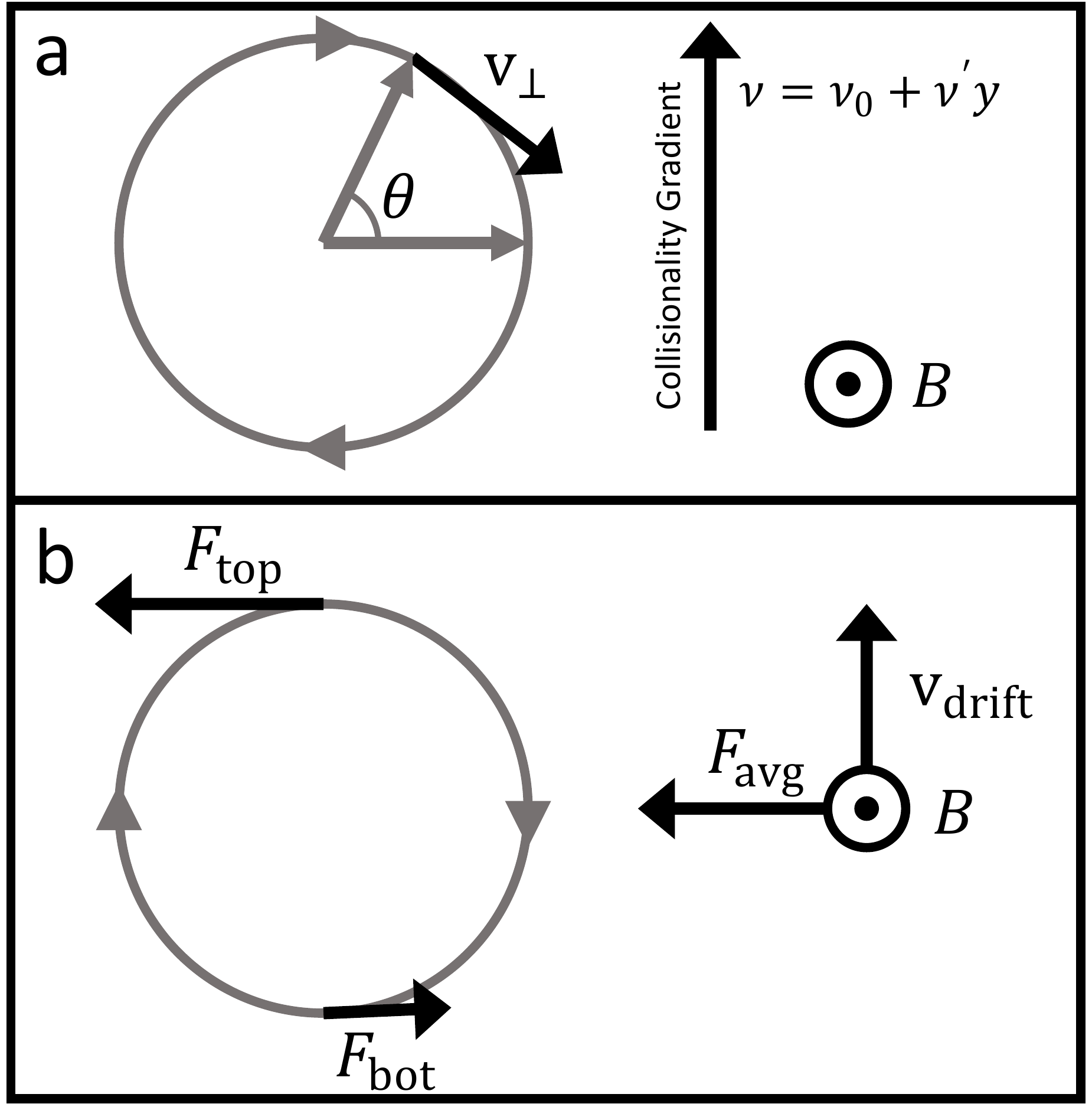}}
	\caption{Heuristic description of the collision gradient drift.
	(a) Problem setup. 
	An ion on a gyro-orbit with velocity $\vp$ in a magnetic field $\Bvec \parallel \hat{z}$ experiences slowing-down collisions at a rate $\nu_s = \nu_0 + \nu' y$.
	(b) The slowing force is greater in regions of higher collisionality, resulting in a net force in the $-\hat{x}$ direction, and thus an $\mathbf{F}\times \Bvec$ drift in the $\hat{y}$ direction, parallel to the collision gradient.}
	\label{fig:heuristic_drift}
\end{figure}

First, consider an ion gyrating with speed $\vp$ around a $\hat{z}$-directed magnetic field, in the presence of a collisionality gradient $\nabla \nu_s \parallel \hat{y}$ (Fig. \ref{fig:heuristic_drift}a).
We can write 
\beq
	\nu_s(y) = \nu_0 + \nu' y = \nu_0 + \nu' \lp \frac{\vp}{\Omega} \sin \theta \rp.
\eeq
Neglecting for the moment diffusive terms in the collision Langevin equation, the particle experiences a deterministic force
\beq
	\mathbf{F}_c = -m_i \nu_s(y) \vvec = -m_i \lp \nu_0 + \nu' \frac{\vp}{\Omega} \sin \theta \rp \begin{pmatrix} \vp \sin \theta\\ -\vp \cos \theta \end{pmatrix},
\eeq
where $m_i$ is the ion mass and $\Omega$ is the ion gyrofrequency.
Averaging this force over $\theta$, we find
\beq
	\langle \mathbf{F}_c \rangle_\theta= -\frac{m_i}{2} \frac{\vp^2}{\Omega} \nu' \hat{x}.
\eeq
Our $\mathbf{F}_c \times \Bvec$ drift is thus given by
\beq
	\vvec_{d,c} = \frac{\langle \mathbf{F} \rangle_\theta \times \hat{z}}{q B} = \frac{1}{2} \frac{\vp^2}{\Omega^2} \nu' \hat{y} = \frac{1}{2} \rho^2 \nu' \hat{y} = \frac{1}{2} \rho^2 \nabla \nu_s . \label{eq:heuristic_drift}
\eeq
Thus the slowing down collisions draw the ions into regions of higher collisionality (Fig. \ref{fig:heuristic_drift}b).

Now consider that the background might have a flow velocity.
Then our force at any point in the orbit is supplemented by
\begin{align}
	\mathbf{F}_\text{flow} &= m_i \nu_s(y) \vbgvec \\
	&= m_i \lp \nu_0 + \nu' \frac{\vp}{\Omega} \sin \theta \rp  \begin{pmatrix} v_{bgx} \\ v_{bgy} \end{pmatrix},
\end{align}
where $m_i$ is the ion mass and $\Omega$ is the ion gyrofrequency.
Averaging this force over $\theta$, we find
\beq
	\langle \mathbf{F}_\text{flow} \rangle_\theta= -\frac{m_i}{2} \frac{\vp^2}{\Omega} \nu' \hat{x} + m_i \nu_0 \vbgvec.
\eeq
Our $\mathbf{F}_\text{flow} \times \Bvec$ drift is thus given by
\begin{align}
	\vvec_{d,\text{flow}} &= \frac{\langle \mathbf{F}_\text{flow} \rangle_\theta \times \hat{z}}{q B} \\
	&=  \frac{m_i \nu_0}{qB} \vbgvec \times \hat{z} \\
	&= \frac{\nu_0}{\Omega} \vbgvec \times \hat{z} \\
	&= \frac{\langle \nu_s \rangle}{\Omega} \vbgvec \times \hat{b}. 
\end{align}
Thus our total $F \times B$ drift due to interactions with the background ions is
\beq
	\vvec_d = \frac{1}{2} \rho^2 \nabla \nu_s  + \frac{\langle \nu_s \rangle}{\Omega} \vbgvec \times \hat{b}. \label{eq:heuristic_drift}
\eeq

\section{Random walk derivation of mass-based separation}

We can also derive our results for energy dissipation and probability of exiting at each side of the device, in the constant-$\eps$ case, by considering the random walk of a single ion across the device.

Consider a system of length $2L$ parameterized by $-L<x<L$, perpendicular to a constant magnetic field $B$, with a buffer density given by $n_b \propto e^{x/L_c}$.
We will assume that circularly-polarized ion cyclotron heating maintains the desired species at an energy $\eps_\perp \gg T_j$, where $j$ is the buffer species, so that energy is added to the particles without itself inducing any drift.
These superthermal ions will be more susceptible to the biased drift, and thus will exit preferentially on one side of the device, while light and thermal ions will tend to exit equally often at either side of the device.

We will work to zeroth order in the quantity $\nu/\Omega$.
Thus we will have 
\beq
	\frac{v_d}{D_\perp} = \frac{\tfrac{1}{2}\rho^2 \nu_s'}{\tfrac{1}{2}\rho^2 \nu_D} = \frac{\eta}{L_c} . \label{eq:vd_D}
\eeq
This will simplify the calculation dramatically by ensuring that the probabilities are constant in the random walk across the length of the device.

Ions enter the device along the magnetic field line at $x = 0$; we need to determine the relative probabilities of exit at $x=\pm L$.
We will begin by recasting the problem as a discrete random walk, for which these probabilities are easily calculated, and then take the continuous limit.
Thus consider a biased random walk, with space step $\Delta x$ and time step $\Delta t$, with probabilities $p$ and $q=1-p$ of stepping toward $+x$ and $-x$ respectively.
Our drift velocity is then given by 
\beq
	v_d = \frac{\langle x(t+\Delta t) \rangle - \langle x(t) \rangle}{\Delta t} = (2p - 1) \frac{\Delta x}{\Delta t}. \label{vd_random}
\eeq
We can eliminate $\Delta t$ by noting that $D_\perp = \Delta x^2 / 2\Delta t$.
Thus $p$ is given by
\beq
	p = \frac{\eta}{4} \Delta x + \frac{1}{2}.
\eeq
Now we can take advantage of our standard random walk results.
In a system consisting of $2N$ steps, a biased random walk beginning in the middle will exit on the $+x$ side with probability
\beq
	P_+ = \frac{1-(q/p)^N}{1-(q/p)^{2N}}. \label{eq:Pp_rw}
\eeq
By invoking equation (\ref{eq:vd_D}), noting that $N = L/\Delta x$, and taking the limit as $\Delta x\rightarrow 0$, we find
\beq
	P(+x) = \frac{1}{1+\exp[-\eta L/L_c]}. \label{eq:p_sym}
\eeq
Thus separation increases with the slowing to scattering ratio $\eta$, with the system size $L$, and with the collisionality gradient $L_c^{-1}$.

We can also calculate the energy dissipated per ion.
The average number of steps $S$ taken in biased random walk is given by
\beq
	T= \frac{N}{p-q} \frac{(p/q)^N - 1}{(p/q)^N + 1}. \label{eq:T_rw}
\eeq
Following our calculation above, this becomes
\beq
	T = \frac{2}{\eta} \frac{1-\exp[-\eta L/L_c] }{1+\exp[-\eta L/L_c]} \frac{L L_c}{\Delta x^2}.
\eeq
We need to relate this to the energy dissipated per step.
The power dissipated is given by
\beq
	P_{dis} = \eps_\perp \lp 2\nu_s - \frac{1}{2}\nu_\perp - \nu_\parallel \rp = 2\eps_\perp \nu_D \lp  \eta - 1 \rp
\eeq

Thus the energy $\Delta \eps_\perp$ dissipated per step is given by (making use of Eq. (\ref{eq:dperp}))
\beq
	\Delta \eps_\perp = P_{dis} \Delta t = P_{dis} \frac{\Delta x^2}{2D} = \frac{2 \eps_\perp}{\rho^2} \lp  \eta - 1 \rp \Delta x^2.
\eeq
The total energy per particle $E$ dissipated is given by the product of $\Delta \eps_\perp$ and $T$, and so we find
\beq
	E = 4 \eps_\perp \frac{L L_c}{\rho^2} \frac{\eta-1}{\eta} \lp \frac{1-\exp[-\eta L/L_c] }{1+\exp[-\eta L/L_c]} \rp . \label{eq:e_sym}
\eeq

\subsubsection{Model Extension}

We can extend this model to a system where the particles are not introduced precisely between the two sinks; rather the system, of total size $L_s$, extends from $-L_- < x < L_s-L_-$, with $N$ steps to the left and $M$ steps to the right.
The generalizations of the random-walk results (Eqs. \ref{eq:Pp_rw} and \ref{eq:T_rw}) to this new scenario are
\beq
	P_+ = \frac{1-(q/p)^N}{1-(q/p)^{N+M}},
\eeq
\beq
	\frac{M - (q/p)^N \lp M+N \rp +N(q/p)^{N+M}}{(p/q) \lp 1-(q/p)^{M+N} \rp}.
\eeq
The generalization of Eqs (\ref{eq:p_sym}) and (\ref{eq:e_sym}) to this new scenario are, taking $N = L_-/\Delta x$, $M = (L_s - L_-)/\Delta x$, and letting $\Delta x \rightarrow 0$:
\beq
	P_+ = \frac{1-\exp[-\eta L_-/L_c]}{1-\exp[-\eta L_s/L_c]},
\eeq
\beq
	T = \frac{2}{\eta} \frac{L_s (1- e^{-\eta L_-/L_c}) - L_-(1- e^{-\eta L_s/L_c})}{1- e^{-\eta L_s/L_c}} \frac{L_c}{\Delta x^2},
\eeq
and
\begin{align}
	E &= 4 \eps_\perp \frac{L_c}{\rho^2} \frac{\eta-1}{\eta} \notag\\
	& \qquad \times \lp L_s  \frac{1- e^{-\eta L_-/L_c}}{1- e^{-\eta L_s/L_c}} -L_- \rp.
\end{align}
These formula are valid for all species present; each species will differ in its value of $\eta$.

\section{Thermal Force} \label{sec:thermalForce}

In our original dynamic equation (Eq. \ref{eq:ionDynamics}), from which we derived our drift, we do not include a thermal force.
In order to properly treat thermal effects, a full kinetic model would be necessary.
However, it is instructive to see what would happen if we did include this term in a simple fluid model.

As we are studying the slow drift \textit{across} the field lines, where the
fast cyclotron rotation \textit{around} the lines has been averaged out, we
consider the classical adiabatic fluid model for a population of charged
particles with gyrofrequency $\Omega$, collision frequency $\nu $%
, fluid velocity $\mathbf{v}$ and pressure $p$:
\begin{equation}
\Omega \mathbf{v}\times \mathbf{b}-\nu \mathbf{v}-\frac{\mathbf{\nabla }p%
}{mn}=0, \label{eq:coldFluid}
\end{equation}
where $m$ is the particle mass and $n$ the density of particles, $\mathbf{b}$ is a
unit vector along the magnetic field.
The entropic momentum exchange term $\mathbf{\nabla }p$ can be expanded as $%
k_{B}T\mathbf{\nabla }n+k_{B}n\mathbf{\nabla }T$, where $T$ is the ion
temperature.

 Considering the $k_{B}T\mathbf{\nabla }n$ term alone allows to
describe the particles (Braginskii) diffusion when Eq. (\ref{eq:coldFluid}) is solved for $\mathbf{v}$. 
This is the usual simple fluid derivation of magnetized diffusion.
Thus
\beq
\left( \frac{\Omega}{\nu }\right) n\mathbf{v}\times \mathbf{b}-n\mathbf{%
v} =\left( \frac{k_{B}T}{m\nu }\right) \mathbf{\nabla }n, \\
\eeq
becomes
\begin{align}
	n\mathbf{v} & =-\frac{\left( k_{B}T/m\nu \right) }{1+\frac{\Omega^{2}}{\nu
^{2}}}\mathbf{\nabla }n-\frac{\left( k_{B}T/m\nu \right) }{1+\frac{\nu ^{2}}{\Omega^{2}}}%
\left( \mathbf{b\cdot \nabla }n\right) \mathbf{b} \notag\\
	& \qquad -\frac{\Omega}{\nu }\frac{\left( k_{B}T/m\nu
\right) }{1+\frac{\Omega^{2}}{\nu ^{2}}}\mathbf{\nabla }n\times \mathbf{b},
\end{align}
where we recognize the classical unmagnetized diffusion coefficient $%
k_{B}T/m\nu $ and the 3 components of the flux $n\mathbf{v}$, i.e. diffusive flux across and along the density gradient, and the diamagnetic drift.
The parallel diffusion term is absent in our model, and so the only new contribution is a diamagnetic drift, which is not a gyrocenter drift but rather a circulating fluid flow.

Now considering the $k_{B}n\mathbf{\nabla }T$ term alone allows us to describe the thermal force when Eq. (\ref{eq:coldFluid}) is solved for $\mathbf{v}$.
Thus
\beq
\left( \frac{\Omega}{\nu }\right) \mathbf{v}\times \mathbf{b}-\mathbf{v}
=\left( \frac{k_{B}T}{m\nu }\right) \frac{\mathbf{\nabla }T}{T}, \\
\eeq
becomes
\begin{align}
	\mathbf{v}&=-\frac{\left( k_{B}T/m\nu \right) }{1+\frac{\Omega^{2}}{\nu
^{2}}}\frac{\mathbf{\nabla }T}{T} 
	-\frac{\left( k_{B}T/m\nu \right) }{1+\frac{%
\nu ^{2}}{\Omega^{2}}}\left( \mathbf{b\cdot }\frac{\mathbf{\nabla }T}{T}%
\right) \mathbf{b}\notag\\
	& \qquad -\frac{\Omega}{\nu }\frac{\left(
k_{B}T/m\nu \right) }{1+\frac{\Omega^{2}}{\nu ^{2}}}\frac{\mathbf{%
\nabla }T}{T}\times \mathbf{b}.
\end{align}
As we are interested by the drift dynamics across the magnetic lines in the
strongly magnetized regime $\nu \ll \Omega$ we can restrict this
general result to the simple expression: 
\beq
\mathbf{v}_T\approx -\frac{\nu ^{2}}{\Omega^{2}}\left( k_{B}T/m\nu
\right) \frac{\mathbf{\nabla }T}{T}-\frac{\nu }{\Omega}\left(
k_{B}T/m\nu \right) \frac{\mathbf{\nabla }T}{T}\times \mathbf{b}.
\eeq
The second term is just the thermal diamagnetic fluid flow
resulting from a temperature gradient, while the first one is the \textit{%
thermal force effect} which must be compared to the collisionality gradient
drift effects $\mathbf{v}_{d}$ identified and analyzed in our paper.

The first point to note is that the thermal force effect will, in general, lead to a drift down the temperature gradient.
If we identify our fluid temperature with our minority species energy, this temperature will be lower in regions of higher collisionality when heating is constant across the device.
Thus the thermal force drift will push the particle into regions of higher collisionality, adding to the collisional drift rather than disrupting it.

Now we can calculate the relative magnitude of the two drifts.
Introducing the mean Larmor radius $\rho$,%
\beq
\mathbf{v}_{T}=-\frac{\nu }{\Omega^{2}}\left( k_{B}T/m\right) \frac{%
\mathbf{\nabla }T}{T}\approx -\nu \rho^{2}\frac{\mathbf{\nabla }T}{T} .
\eeq
This must be compared to : 
\beq
\mathbf{v}_{d}\approx \frac{\nu \rho^{2}}{2}\frac{\mathbf{\nabla }\nu }{%
\nu } 
\eeq
Thus, the criterion to neglect the thermal force is: 
\beq
\frac{\mathbf{\nabla }\nu }{\nu }>\frac{\mathbf{\nabla }T}{T} .
\eeq
This condition was satisfied for our first, simple model with uniform minority energy, but not for our realistic heating model (Section \ref{sec:realisticHeating}), where energy scaled as background density (and thus collisionality) to the negative second power (Eqs. \ref{eq:eps_heat}-\ref{eq:eps0}).
However, it is of the same approximate magnitude as the collisional drift, and as mentioned earlier, points in the same direction.
Thus the thermal effects are not likely to substantially change the conclusions of the paper.

It should be noted that here we have employed the simplest possible fluid closure, which does not allow a heat flux; then we demanded a specific temperature gradient, without specifying how it was applied, or even how temperature should be consistently defined; and finally, demanded a steady state.
Thus, this drift is simply the consistent drift with the conditions we have, perhaps unphysically, demanded.
In order to more accurately determine what thermally-related drifts will result when we heat a minority ion population, one should properly consider heat fluxes and heat exchange between species, as well as kinetic effects such as thermo- and baro-diffusion which link inhomogeneities in temperature to additional fluxes.
Such a problem is outside the scope of this paper, and so we limit our considerations to the collisional drift.

\section*{References}
\bibliography{collisions_prop}

\clearpage
\newpage

\end{document}